\documentclass[9pt,conference]{IEEEtran}

\usepackage[preprint]{waspaa25}

\usepackage{bm} 
\usepackage{tcolorbox}
\usepackage{adjustbox}

\definecolor{lightblue}{RGB}{173, 216, 230}   
\definecolor{lightorange}{RGB}{245, 222, 179} 
\definecolor{lightpink}{RGB}{255, 228, 225}   

\definecolor{lightlightgray}{gray}{0.95}

\def\ear{{\includegraphics[width=0.0115\textwidth]{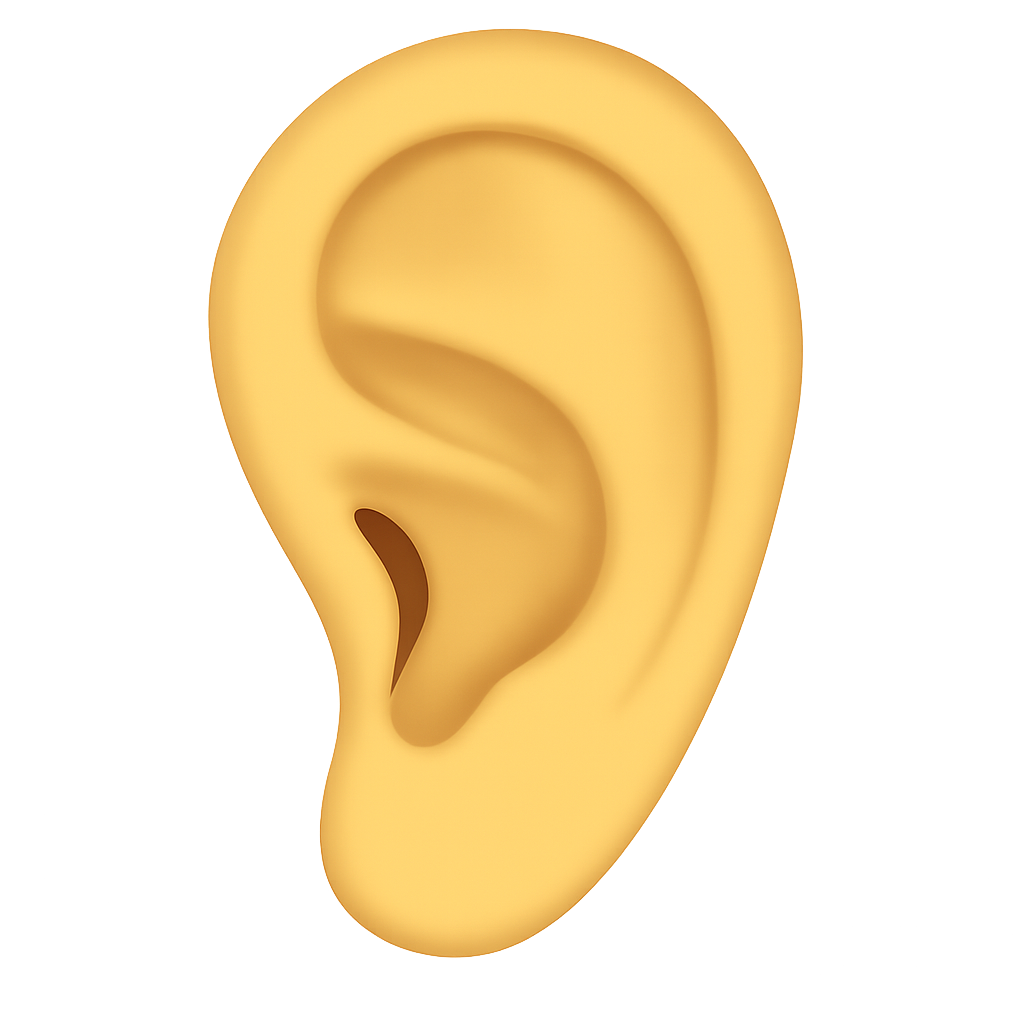}}{}}
\def\eye{{\includegraphics[width=0.013\textwidth]{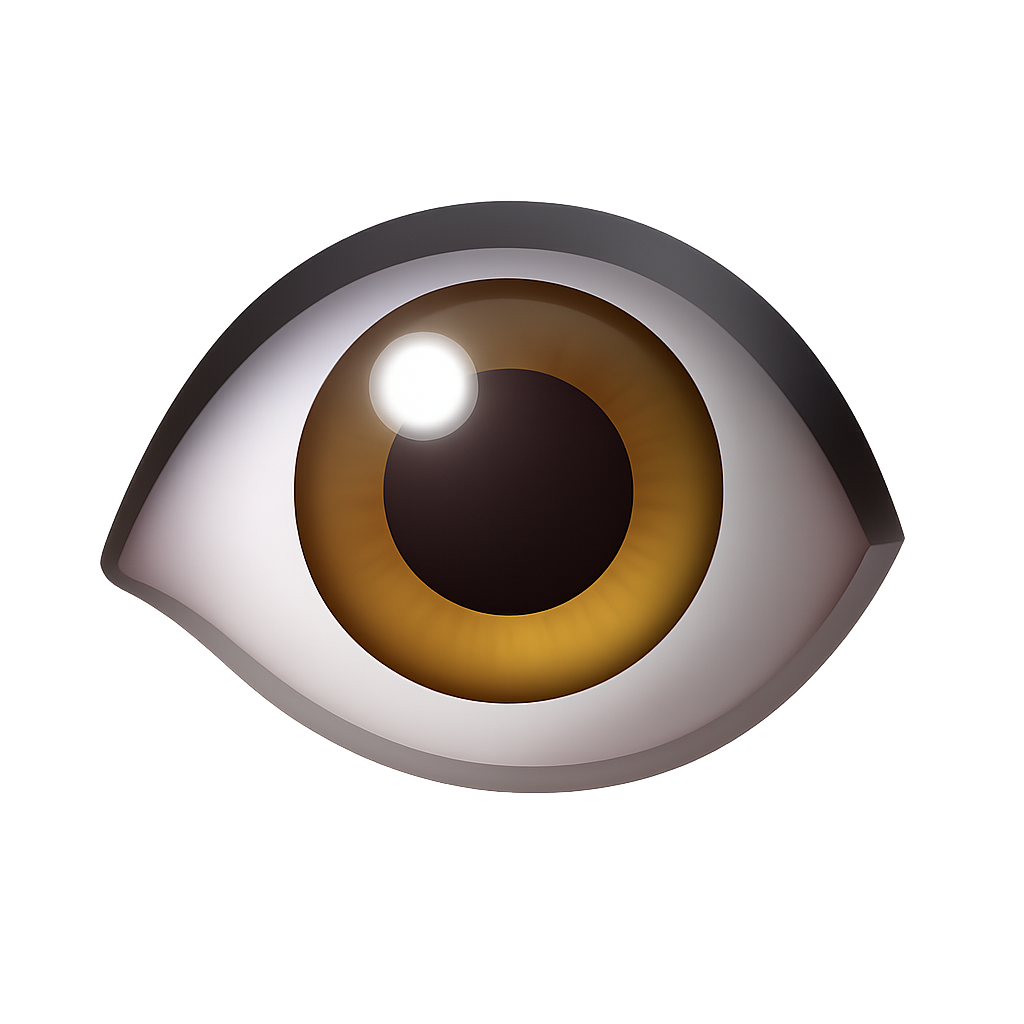}}{}}

\def\lear{{\includegraphics[width=0.015525\textwidth]{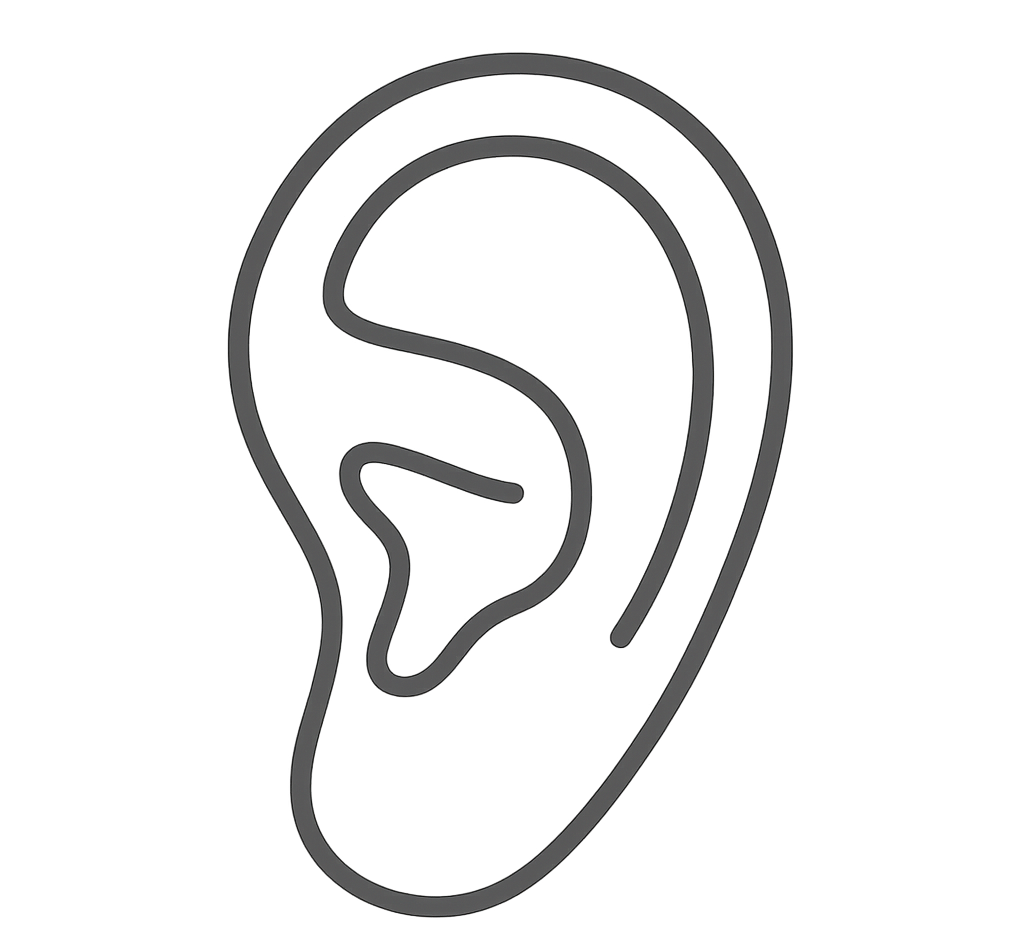}}{}}
\def\leye{{\includegraphics[width=0.01755\textwidth]{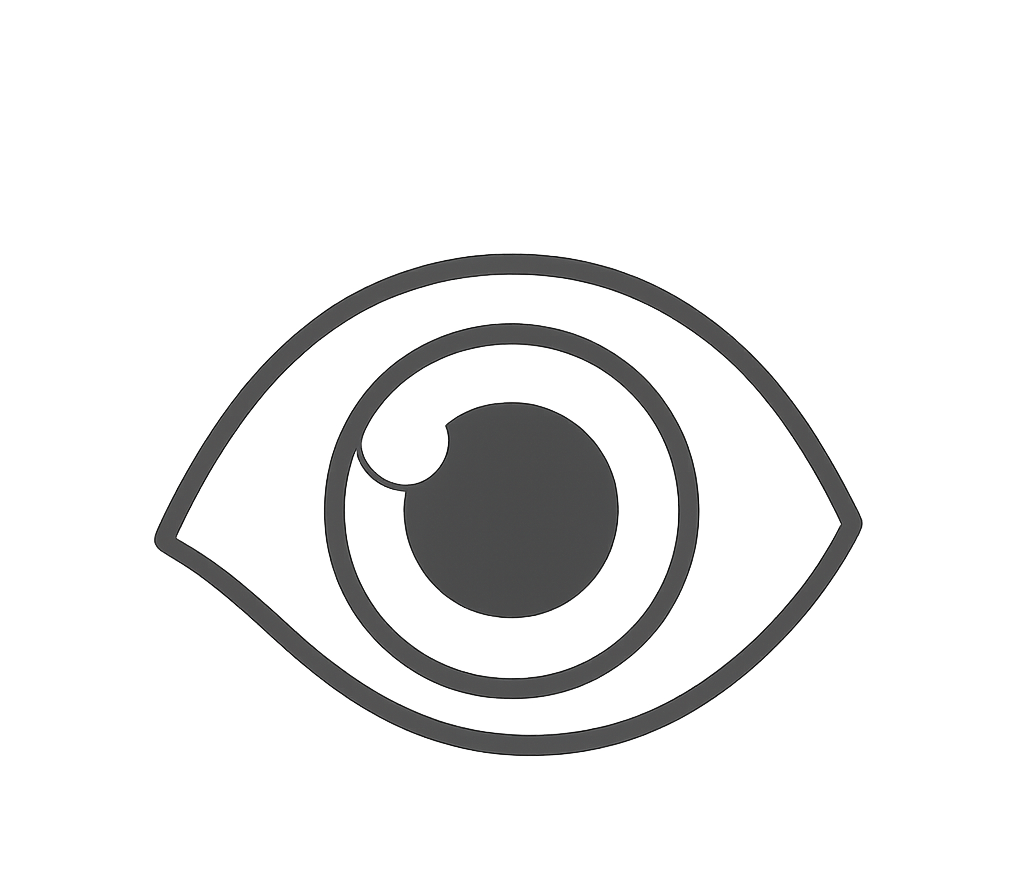}}{}}

\title{Bridging Ears and Eyes: Analyzing Audio and Visual Large Language Models to Humans in Visible Sound Recognition and Reducing Their Sensory Gap via Cross-Modal Distillation}

\name{Xilin Jiang$^{\ear\sharp}$,
      Junkai Wu$^{\eye\flat}$\thanks{$^{\ear\eye}$The first two authors contributed equally.},
      Vishal Choudhari$^\sharp$,
      Nima Mesgarani$^\sharp$}
\address{$^{\sharp}$Columbia University, NY, USA \;
$^{\flat}$University of Washington, WA, USA
}

\begin{document}

\maketitle

\begin{abstract}
Audio large language models (LLMs) are considered experts at recognizing sound objects, yet their performance relative to LLMs in other sensory modalities, such as visual or audio-visual LLMs, and to humans using their ears, eyes, or both remains unexplored. To investigate this, we systematically evaluate audio, visual, and audio-visual LLMs,  specifically Qwen2-Audio, Qwen2-VL, and Qwen2.5-Omni, against humans in recognizing sound objects of different classes from audio-only, silent video, or sounded video inputs. We uncover a performance gap between Qwen2-Audio and Qwen2-VL that parallels the sensory discrepancy between human ears and eyes. To reduce this gap, we introduce a cross-modal distillation framework, where an LLM in one modality serves as the teacher and another as the student, with knowledge transfer in sound classes predicted as more challenging to the student by a heuristic model. Distillation in both directions, from Qwen2-VL to Qwen2-Audio and vice versa, leads to notable improvements, particularly in challenging classes. This work highlights the sensory gap in LLMs from a human-aligned perspective and proposes a principled approach to enhancing modality-specific perception in multimodal LLMs.

\end{abstract}

\section{Introduction}
\label{sec:intro}

Humans perceive the world through multiple sensory systems, with the auditory and visual systems playing dominant roles. Each system is uniquely specialized: the auditory pathway is sensitive to temporal cues and fine-grained frequency patterns \cite{romanski2009primate}, while the visual pathway is attuned to spatial structure and motion \cite{ungerleider2000mechanisms}. To the brain, audio stimuli travel from the cochlea to the auditory cortex, and visual stimuli travel from the retina to the visual cortex. These distinct routes reflect fundamental differences in how each modality encodes and interprets the environment. Yet, the two systems are interconnected at higher cortical levels, enabling multi-sensory integration for robust perception, such as recognizing a barking dog from its shape or sound.

Recent advances in multimodal large language models (LLMs) \cite{latif2023sparks, zhang2024vision} have introduced architectures that parallel human sensory systems. Audio and visual LLMs typically begin with a modality-specific encoder, such as Whisper \cite{whisper} for audio or Vision Transformer (ViT) \cite{vit} for images, that transforms continuous sensory inputs into contextual representations. These representations, along with textual prompts, are then processed by a shared Transformer decoder-only language model \cite{gpt1, gpt2}. This unified model operates analogously to the association cortices of the brain, where signals from different sensory pathways are integrated to support higher-level reasoning.

Despite strong performance, the sensory boundaries and complementarity of audio and visual LLMs remain unexplored, especially in direct comparison to identical acoustic scenes. Studies in neuroscience and psychology show that human sensory systems exhibit modality-specific biases: audition excels in dynamic or occluded environments, while vision dominates in spatially rich scenes \cite{ernst2004merging, shams2008benefits}. We ask \textit{whether a sensory gap exists in LLMs} and, if so, \textit{whether it mirrors that observed in humans.} Once we discover this modality-specific performance gap between audio and visual LLMs, another natural question arises: \textit{can we reduce the sensory gap by leveraging another modality?} This question is inspired by the \textit{synesthesia} phenomenon in humans \cite{ramachandran2001synaesthesia, cytowic2002synesthesia}, where stimulation of one sensory can evoke perceptual experiences in another—suggesting the possibility of cross-sensory reinforcement. 

To answer both questions, this study is structured into two parts: \textbf{\textit{analysis}} and \textbf{\textit{distillation}}. In the analysis part, we evaluate audio LLM Qwen2-Audio \cite{qwen2audio}, visual LLM Qwen2-VL \cite{qwen2vl}, and audio-visual LLM Qwen2.5-Omni \cite{qwen2.5omni}, among other multimodal LLMs and real humans, in the audio-visual VGGSound dataset \cite{vggsound}. Each model receives audio-only, silent video, or both and is prompted to classify the sounding object from available sensors. This setup can reveal modality-specific strengths and weaknesses and eventually shows that a sensory gap exists for LLMs and aligns with this for humans, with some classes where Qwen2-Audio still underperforms human listeners. In the distillation part, we propose a cross-modal framework to distill knowledge between LLMs. A \textit{heuristic switch}, trained on our analysis findings, routes supervision from the stronger to the weaker modality. Finally, our distilled Qwen2-Audio achieves over 20\% accuracy gain, surpassing its visual teacher and matching the audio-visual Qwen2.5-Omni.

In summary, our study highlights two key insights: first, audio and visual LLMs exhibit a sensory gap that mirrors that of humans, and second, these gaps can be effectively reduced through targeted cross-modal knowledge distillation. Together, these findings offer a deeper understanding of modality specialization in LLMs and a principled framework for enhancing their perceptual capabilities.

\begin{figure*}[t]
  \centering
\centerline{\includegraphics[width=\linewidth]{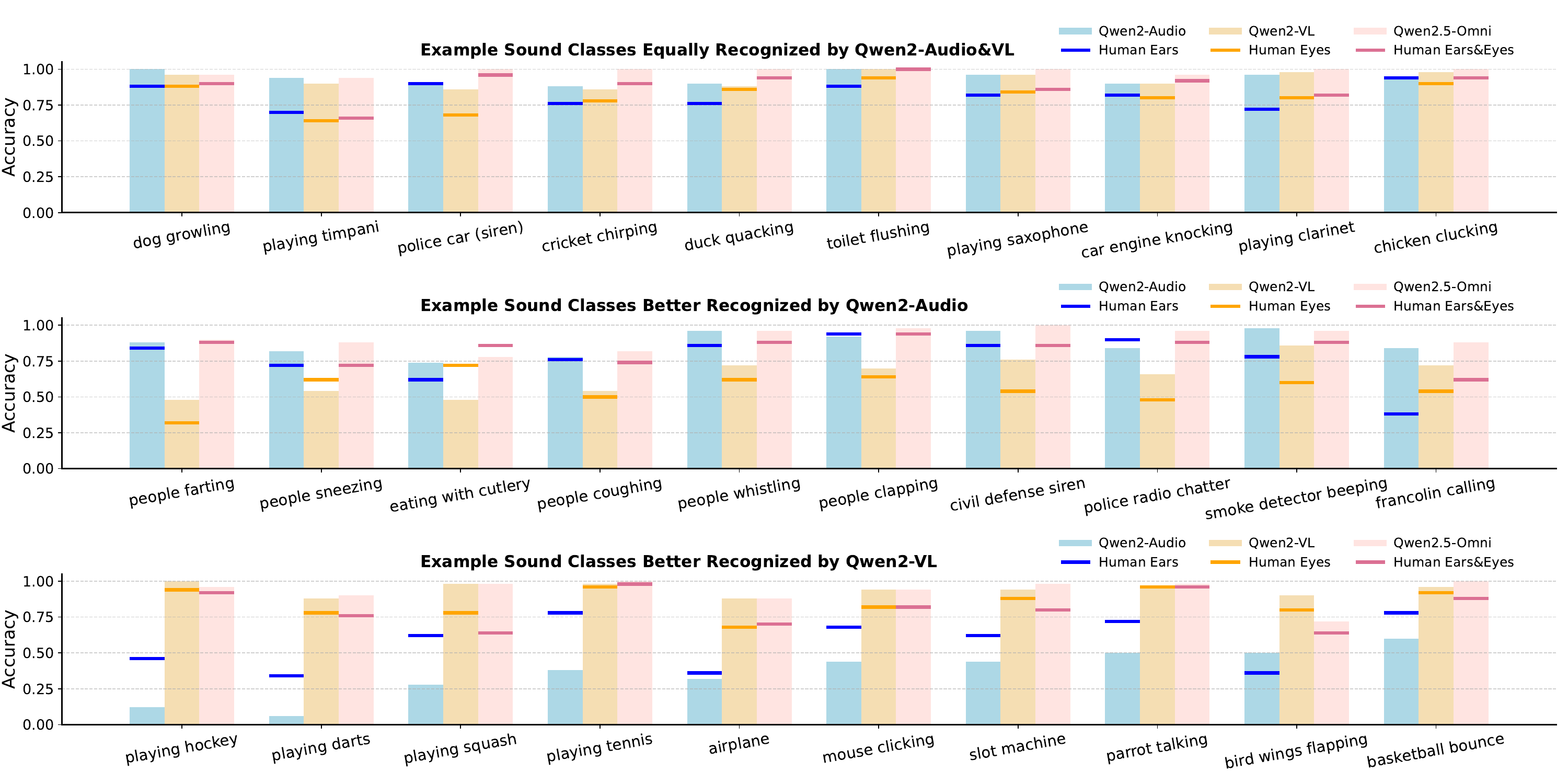}}
  \caption{We compared Qwen2-Audio, Qwen2-VL, and Qwen2.5-Omni among themselves and against humans on 30 classes $\times$ 50 samples/class from VGGSound analysis set under three input conditions of the same acoustic scene: audio-only, silent video, and sounded video. Class-wise analysis reveals that the sensory gap among LLMs (audio vs. visual vs. both) mirrors human perception differences between ears, eyes, and their integration. The overall accuracy of 30 classes shown in this figure is: Qwen2.5-Omni (94.4\%) $>$ Human Ears\&Eyes (84.2\%) $>$ Qwen2-VL (83.9\%) $>$ Human Eyes (74.1\%) $>$ Qwen2-Audio (72.5\%) $>$ Human Ears (71.9\%). Despite slightly higher overall accuracy, Qwen2-Audio underperforms human listeners in most challenging sound classes (bottom row).}
  \label{fig:ears_vs_eyes}
\end{figure*}

\section{Related Works}
\textbf{(Visible) sound recognition} has been studied across both audio and visual modalities. Models such as PANN \cite{pann} and AST \cite{ast} for audio, and ViT \cite{vit} and SAM \cite{sam} for images have set state-of-the-art results. More recent methods like CLIP \cite{clip} and CLAP \cite{clap} integrate language models for open-vocabulary recognition, producing free-form captions rather than selecting from a pre-defined label set.

\noindent
\textbf{Audio and Visual LLMs} represent the latest evolution in sound recognition, while their capabilities extend to broader scene analysis and reasoning tasks \cite{latif2023sparks, zhang2024vision}. They typically integrate an audio or visual encoder, like those mentioned above, with a pretrained LLM.
While prior studies have evaluated audio or visual LLMs individually \cite{audiobench, lee2024vhelm}, and a few for audio-visual LLMs \cite{avhbench, avtrustbench}, to our knowledge, this is the first work to directly contrast audio and visual LLMs and humans on the same acoustic scenes.

\noindent
\textbf{Knowledge Distillation for LLMs} typically happens between two text-only LLMs or two multimodal LLMs of the same modalities \cite{gu2024minillm, efficientvlm}. Only a few non-LLM works explore cross-modal distillation between audio and video models \cite{chen2021distilling, sarkar2024xkd}.

\section{Cross-Modal Analysis}
\label{sec:analysis}

\subsection{Multimodal LLMs and Data}

This study focuses on \textit{single-sensory} multimodal LLMs that perceive the environment with either the auditory or visual modality, but not both. To this end, our research subjects are three state-of-the-art audio LLMs: Qwen2-Audio \cite{qwen2audio}, Qwen-Audio \cite{qwenaudio}, and Kimi-Audio \cite{kimiaudio}, and three state-of-the-art visual LLMs: Qwen2-VL \cite{qwen2vl}, Qwen2.5-VL \cite{qwen2.5vl}, and VideoLLaMA3 \cite{videollama3}. Our analysis places particular emphasis on Qwen2-Audio and Qwen2-VL, as they both share Qwen2 \cite{qwen2} as the language model, allowing us to isolate and investigate the impact of sensory modality in model behavior. This comparison is analogous to a human blocking either their ears or eyes—a scenario we also replicated and examined in human evaluation. To contextualize the limitations of single-sensory perception, we also evaluated an audio-visual LLM Qwen2.5-Omni \cite{qwen2.5omni}, which processes both modalities jointly and serves as an upper bound for multimodal understanding.

VGGSound \cite{vggsound}, a large-scale audio-visual dataset consisting of in-the-wild videos spanning a wide range of everyday sound events, serves as an ideal benchmark for evaluating and comparing multimodal systems. After removing corrupted videos and cleaning the labels, we curated 149,460 samples for finetuing LLMs (used in the next section), 14,300 for analysis (used in this section), and 14,202 for testing. The analysis set is randomly excluded from the training set, while the test set corresponds to the original split provided by VGGSound. All sets contain 10-second videos of one of 286 sound classes, including animal vocalizations, musical instruments, and various artificial sounds.

We evaluated all LLMs and human participants using multiple-choice sound recognition questions. In preliminary tests, we unfortunately observed that neither LLMs nor humans were able to reliably identify the correct sound label when no candidate choices were provided. The question prompt used in this study is shown below:

\begin{tcolorbox}[colback=gray!3!white, colframe=gray!75!black, title={Shared Prompt to Audio, Visual, and Audio-Visual \\ LLMs, as well as Humans}]
\small
\texttt{<audio.wav>}, \texttt{<silent\_video.mp4>} \\ or \texttt{<sounded\_video.mp4>} \\
Classify the sounding object into one of the categories below:
\begin{itemize}
    \item[A.] police car
    \item[B.] ice cream van
    \item[C.] (omitted...)
    \item[J.] plastic bottle crushing
\end{itemize}
Respond only with a single letter from A to J.
\end{tcolorbox}

Depending on the model’s sensory channel and the experimental setting, each LLM or human participant was presented with either an audio clip, a silent video, or a sounded video. The multiple-choice options were sampled from the dataset according to the frequency of label occurrences, and the ground-truth label was randomly positioned among them to mitigate positional bias. The \textbf{same} prompt and options were used for all LLMs and human participants to ensure a fair and consistent evaluation protocol. Finally, all audio clips were converted to 16 kHz mono-channel format, and all videos were downsampled to 1 fps with a maximum resolution of 262,144 (512$\times$512, while preserving the aspect ratio) pixels due to GPU memory constraint.

\begin{table}[t]
\centering
\caption{Average multiple-choice accuracy in the VGGSound analysis and test sets (286 classes $\times$ $\sim$50 samples/class). Our distilled models significantly outperform the original ones and other multimodal LLMs.}
\label{tab:main}
\begin{adjustbox}{width=0.8\linewidth,center}
\begin{tabular}{lcc}
\toprule
\textbf{Multimodal LLM} & \textbf{Analysis Set} & \textbf{Test Set} \\
\hline
\rowcolor{lightblue}
\multicolumn{1}{l}{
\textbf{\textcolor{gray}{Audio LLM}} \lear} & & \\
\hline
Qwen2-Audio-7B-Instruct \cite{qwen2audio} & 71.9 & 69.0 \\
- \textbf{distilled from Qwen2-VL} & \textbf{92.6} & \textbf{89.4} \\
Qwen-Audio-Chat (7B) \cite{qwenaudio} & 63.2 & 60.4 \\
Kimi-Audio-7B-Instruct \cite{kimiaudio} & 70.6 & 66.7 \\
\hline
\rowcolor{lightorange}
\multicolumn{1}{l}{ \textbf{\textcolor{gray}{Visual LLM}} \leye} & & \\ \hline
Qwen2-VL-7B-Instruct \cite{qwen2vl} & 88.5  & 80.8 \\
- \textbf{distilled from Qwen2-Audio} & \textbf{91.4} & \textbf{83.9} \\
Qwen2.5-VL-7B-Instruct \cite{qwen2.5vl} & 88.8 & 81.0 \\
VideoLLaMA 3 \cite{videollama3} & 83.4 & 76.4  \\
\hline
\rowcolor{lightpink}
\multicolumn{1}{l}{
\textbf{\textcolor{gray}{Audio-Visual LLM}} \lear\leye} & & \\ \hline
Qwen2.5-Omni \cite{qwen2.5omni} & 92.6 & 89.7 \\
\bottomrule
\end{tabular}
\end{adjustbox}
\end{table}

\subsection{Performance Analysis of Original LLMs}

\textbf{Overall performance} in the analysis set is shown in Table~\ref{tab:main} to compare sound recognition accuracy across multimodal LLMs. The audio-visual model Qwen2.5-Omni achieves the highest accuracy at 92.6\%, serving as a strong reference point. Among the single-sensory models, visual LLMs consistently outperform their audio counterparts: Qwen2-VL and Qwen2.5-VL reach 88.5\% and 88.8\% accuracy, respectively, while Qwen2-Audio lags behind at 71.9\% and other models are worse. While VGGSound is primarily used in the sound recognition literature for audio models, our results reveal that its visual modality is highly, if not more, informative.

\noindent \textbf{Class-wise analysis} reveals a pronounced sensory gap across different acoustic scenes for multimodal LLMs.  We present $3\times10$ representative classes where Qwen2-Audio performs on par with, better than, or worse than Qwen2-VL in Fig. \ref{fig:ears_vs_eyes}. While Qwen2-VL achieves higher accuracy in 226 out of 286 classes, Qwen2-Audio excels in recognizing human-related actions such as coughing, sneezing, and whistling, where visual cues are limited. In contrast, Qwen2-VL performs better in visually distinctive categories like sports and, interestingly, in cases like parrot sounds and bird wings flapping, which are ambiguous to classify based on sound. Both models show strong and comparable performance on musical instruments and animal vocalizations. Audio-visual Qwen2.5-Omni outperforms or at least performs on par with Qwen2-Audio and Qwen2-VL in nearly all classes.
 
\noindent \textbf{LLMs vs. Humans} comparisons in Fig. \ref{fig:ears_vs_eyes} show that the sensory gap exhibited by LLMs closely mirrors what is observed in humans. We recruited 100 human participants to collectively classify 30 classes, each with 50 samples, under audio-only, silent video, and sounded video conditions—resulting in a total of 4,500 trials (45 trials per participant). Note that the 30 classes were selected solely according to LLMs' performance, without any knowledge of humans' performance. Overall, LLM Qwen2-Audio, Qwen2-VL, and Qwen2.5-Omni outperform humans in each input condition. More importantly, 8 of the 10 classes better recognized by Qwen2-VL and all 10 classes better recognized by Qwen2-Audio are also more accurately recognized by human eyes and ears, respectively, validating the alignment between LLMs and humans' sensory strengths and weaknesses.

\textit{However}, Qwen2-Audio still shows a considerable performance shortfall to human listeners in 9 out of 10 challenging classes (last row), indicating significant room for improvement. This shortfall is not insurmountable, as humans can recognize them with the exact same audio at the same sampling rate (or video with the same resolution and frame rate). This motivates our exploration of cross-modal distillation, elaborated in the next section.

\begin{figure}[!t]
  \centering
\centerline{\includegraphics[width=\columnwidth]{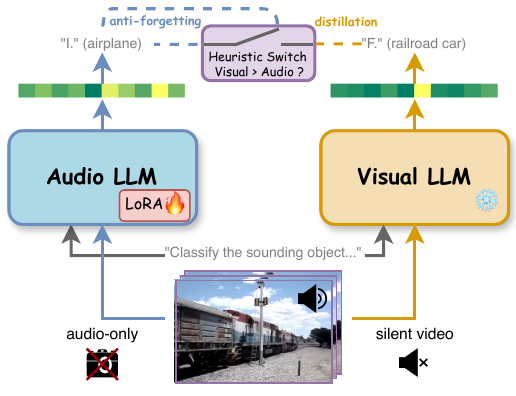}}
  \caption{Proposed cross-modal distillation framework between an audio and a visual LLM. Both LLMs perceive the same acoustic scene but in either audio or video-only format. A heuristic switch predicts whether the student (audio in this figure) LLM underperforms the teacher (visual) LLM for each sample. Based on this decision, the student LLM is finetuned either by the teacher LLM’s output or its own output.}
  \label{fig:distill}
\end{figure}

\section{Cross-modal Distillation}
\label{sec:distillation}

\subsection{Two Modalities, Same Acoustic Scene}

To mitigate the sensory gap revealed in our earlier analysis, we introduce a cross-modal distillation framework, where an LLM from one modality serves as the teacher and an LLM from another modality serves as the student. As illustrated in Fig. \ref{fig:distill}, both models perceive the same acoustic scene; however, one processes it through audio and the other through silent video. In addition, they are both given the same text prompt to classify the sounding object.

The student LLM is trained using the teacher’s predicted labels via cross-entropy loss and/or its output logits via KL divergence, as we will later ablate. This distillation requires no ground-truth label if the teacher is considered reliable. However, as shown in Fig.\ref{fig:ears_vs_eyes}, the teacher (e.g., a visual LLM) does not consistently outperform the student (e.g., an audio LLM) across all classes. To address this, we restrict distillation to samples where the teacher is likely more accurate, as identified by a heuristic model introduced in the following subsection.

\subsection{Heuristic Switch}

The heuristic switch is a binary classifier that predicts whether the teacher LLM of one modality outperforms the student LLM of another modality on a given acoustic scene, specifically, whether Qwen2-VL is more accurate than Qwen2-Audio. When the switch is closed, Qwen2-VL distills knowledge to Qwen2-Audio; when open, distillation is disabled, and Qwen2-Audio is trained on its own outputs to retain prior knowledge and prevent catastrophic forgetting \cite{cl1, cl2}.

This switch is made possible by the sensory gap observed in our analysis. We split the analysis set of 14,300 samples into 12,870 (45 samples/class) for training and 1,430 (5 samples/class) for validation. Based on performance comparisons (like Fig.\ref{fig:ears_vs_eyes} for all 286 classes), we labeled samples from the 226 classes where Qwen2-VL outperforms as 1 (“to distill”) and the remaining 60 classes as 0 (“not to distill”).

We initialized the switch from a pretrained PANN\footnote{\text{Cnn14\_16k\_mAP=0.438.pth}, https://zenodo.org/records/3987831} \cite{pann}, which had not seen VGGSound audio. It was finetuned for 30 epochs using cosine learning rate scheduling with 10\% warmup, a batch size of 32, and an Adam optimizer at a 1e-4 learning rate. The model reached 88.8\% accuracy on the validation set and 89.1\% on the LLM finetuning set. As a proof of concept, this switch is audio-only; future work may explore audio-visual switches for improved decisions.

Notably, the entire cross-modal distillation framework is \textit{mostly unsupervised}. The only supervision required is for training the heuristic switch, which relies solely on the analysis set—less than 10\% of the full LLM finetuning set in this study. Therefore, our framework is easily scalable to in-the-wild data without any labels.

\begin{figure}[!t]
  \centering
\centerline{\includegraphics[width=\columnwidth]{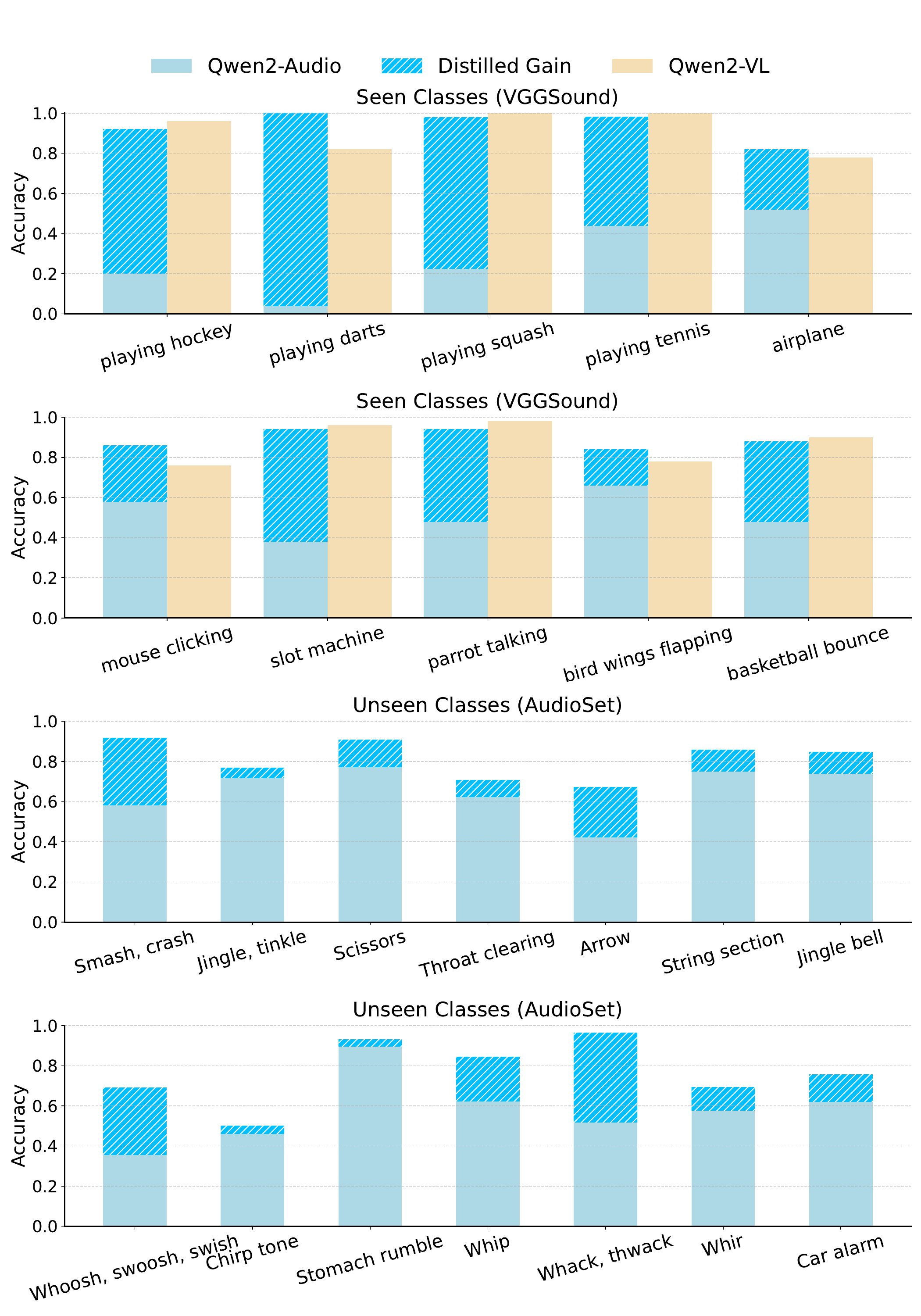}}
  \caption{Class-wise comparison of Qwen2-Audio before and after distillation. Top two rows: The distilled Qwen2-Audio shows substantial accuracy gains on previously challenging classes, comparable to and sometimes surpassing the Qwen2-VL. Bottom two rows: Cross-modal distillation also improves recognizing unseen classes in AudioSet.}
  \label{fig:improve}
\end{figure}

\subsection{Performance Analysis of Distilled LLMs}

We finetuned Qwen2-Audio or Qwen2-VL as the student with the other as the teacher on 149,460 samples (415 hours) from the VGGSound dataset. Both models were finetuned for one epoch using cosine learning rate scheduling with 10\% warmup, an effective batch size of 8 after gradient accumulation, an AdamW optimizer at a 1e-4 learning rate, and LoRA \cite{lora} rank of 16, in a NVIDIA L40 GPU.

\noindent
\textbf{Overall performance} is shown in Table \ref{tab:main}. Cross-modal distillation leads to substantial improvements: Qwen2-Audio increases from 71.9\% to 92.6\% on the analysis set and from 69.0\% to 89.4\% on the test set—surpassing Qwen2-VL and reaching parity with the audio-visual Qwen2.5-Omni. Qwen2-VL also benefits when distilled from Qwen2-Audio, improving to 91.4\% and 83.9\% respectively, though the gains are smaller due to the weaker teacher Qwen2-Audio, better in fewer classes. This shows that our framework brings bidirectional benefits.

\noindent
\textbf{Class-wise gains} are highlighted in Fig.\ref{fig:improve} . Distilled Qwen2-Audio achieves large gains in classes it initially struggled with (corresponding to the last row in  Fig.\ref{fig:ears_vs_eyes}), such as sports, ``parrot talking'' and ``slot machine'', and even suppresses Qwen2-VL in some classes. It also generalizes well to unseen AudioSet \cite{audioset} classes, where we evaluated the model using multiple-choice questions constructed with both correct answers and distractor options sampled from AudioSet's label set. The overall accuracy on the AudioSet evaluation set increased from 60.0 to 72.3 after distillation.

\noindent
\textbf{Ablation} results are presented in Table \ref{tab:ablation}. Both KL divergence on logits (with a temperature of 2) and cross-entropy on teacher labels enable knowledge transfer, but the latter proves more effective in the sound recognition task, approaching the upper bound achieved with cross-entropy on ground-truth labels. Adding the anti-forgetting loss (cross-entropy on the student’s predicted labels) further improves performance by helping retain previously learned knowledge.

In summary, cross-modal distillation significantly boosts multimodal LLM's performance, with Qwen2-Audio matching or even surpassing Qwen2-VL and Qwen2.5-Omni in previously challenging sound classes and generalizing well to unseen classes—highlighting the effectiveness of our scalable, low-supervision framework.

\vspace{-0.15cm}
\section{Discussion and Limitation}
\label{sec:discussion}

Our cross-modal analysis first reveals a strong correspondence between modality-specific gaps in LLMs and human sensory limitations. While our heuristic-guided distillation effectively enhances both audio and visual LLMs, its broader significance lies in demonstrating that these sensory gaps are not fundamental, as targeted knowledge distillation can effectively bridge them. This highlights the potential of leveraging stronger modalities and human-aligned cues \cite{aadllm, sood2023multimodal} as scaffolding to support learning in weaker modalities.

Nevertheless, this preliminary study is scoped to sound recognition tasks, which restrict the degree of reasoning involved. Several human participants offered insightful reflections after completing their trials: \textit{``I used some context clues and took an educated guess.''}, \textit{``The questions encouraged personal reflection.''}, and \textit{``If there was no sound, I guessed what I would hear based on the video.''} These comments underscore the importance of contextual inference, implicit knowledge, and common sense in human perception, aspects not yet examined in the current evaluation and analysis. Expanding this study to more complex acoustic scene and understanding tasks could expose new limitations and inspire new solutions for multimodal LLMs.

\begin{table}
\centering
\caption{Ablation on cross-modal distillation with Qwen2-VL as the teacher and Qwen2-Audio as the student.}
\label{tab:ablation}
\begin{adjustbox}{width=0.8\linewidth,center}
\begin{tabular}{lcc}
\toprule
\textbf{Setting} & \textbf{Analysis Set} & \textbf{Test Set} \\
\hline
KL on Teacher's Logits & 92.1 & 88.7 \\
KL on Teacher's Logits + Labels & 92.3 & 88.9 \\
Ground-truth Labels (Upper bound) & 96.4 & 92.8 \\
Teacher's Labels (Default) & 92.6 & 89.4 \\
\hspace{0.2cm}- without Anti-forgetting Loss & 92.5 & 89.1 \\
\bottomrule
\end{tabular}
\end{adjustbox}
\vspace{-0.05cm}
\end{table}

\section{Conclusion}
\label{sec:conclusion}

This work systematically investigates the sensory capabilities of audio and visual large language models (LLMs) in visible sound recognition and reveals that they exhibit modality-specific strengths and limitations that mirror human listeners and viewers. Motivated by this sensory gap, we propose a cross-modal distillation framework guided by a heuristic switch that selectively routes supervision from the stronger to the weaker modality. Our experiments demonstrate that this approach significantly improves performance in both directions between audio and video, with Qwen2-Audio surpassing its visual teacher, approaching the performance of the audio-visual model, and generalizing well to unseen sound classes. We hope this study inspires the development of multimodal LLMs that are more perceptually aligned to humans and less prone to modality-specific blind spots.

\section*{Acknowledgement}
We appreciate Professor Mari Ostendorf for suggesting heuristic switch. X.J., V.C, and N.M. thank a fund from the National Institutes of Health (NIH-NIDCD) and a grant from Marie-Josee and Henry R. Kravis.


\bibliographystyle{IEEEtran}
\bibliography{refs25}

\begin{thebibliography}{10}
\providecommand{\url}[1]{#1}
\csname url@samestyle\endcsname
\providecommand{\newblock}{\relax}
\providecommand{\bibinfo}[2]{#2}
\providecommand{\BIBentrySTDinterwordspacing}{\spaceskip=0pt\relax}
\providecommand{\BIBentryALTinterwordstretchfactor}{4}
\providecommand{\BIBentryALTinterwordspacing}{\spaceskip=\fontdimen2\font plus
\BIBentryALTinterwordstretchfactor\fontdimen3\font minus \fontdimen4\font\relax}
\providecommand{\BIBforeignlanguage}[2]{{%
\expandafter\ifx\csname l@#1\endcsname\relax
\typeout{** WARNING: IEEEtran.bst: No hyphenation pattern has been}%
\typeout{** loaded for the language `#1'. Using the pattern for}%
\typeout{** the default language instead.}%
\else
\language=\csname l@#1\endcsname
\fi
#2}}
\providecommand{\BIBdecl}{\relax}
\BIBdecl

\bibitem{romanski2009primate}
L.~M. Romanski and B.~B. Averbeck, ``The primate cortical auditory system and neural representation of conspecific vocalizations,'' \emph{Annual review of neuroscience}, vol.~32, no.~1, pp. 315--346, 2009.

\bibitem{ungerleider2000mechanisms}
S.~K. Ungerleider and L.~G, ``Mechanisms of visual attention in the human cortex,'' \emph{Annual review of neuroscience}, vol.~23, no.~1, pp. 315--341, 2000.

\bibitem{latif2023sparks}
S.~Latif, M.~Shoukat, F.~Shamshad, M.~Usama, Y.~Ren, H.~Cuay{\'a}huitl, W.~Wang, X.~Zhang, R.~Togneri, E.~Cambria \emph{et~al.}, ``Sparks of large audio models: A survey and outlook,'' \emph{arXiv preprint arXiv:2308.12792}, 2023.

\bibitem{zhang2024vision}
J.~Zhang, J.~Huang, S.~Jin, and S.~Lu, ``Vision-language models for vision tasks: A survey,'' \emph{IEEE Transactions on Pattern Analysis and Machine Intelligence}, 2024.

\bibitem{whisper}
A.~Radford, J.~W. Kim, T.~Xu, G.~Brockman, C.~McLeavey, and I.~Sutskever, ``Robust speech recognition via large-scale weak supervision,'' in \emph{International conference on machine learning}.\hskip 1em plus 0.5em minus 0.4em\relax PMLR, 2023, pp. 28\,492--28\,518.

\bibitem{vit}
A.~Dosovitskiy, L.~Beyer, A.~Kolesnikov, D.~Weissenborn, X.~Zhai, T.~Unterthiner, M.~Dehghani, M.~Minderer, G.~Heigold, S.~Gelly, J.~Uszkoreit, and N.~Houlsby, ``An image is worth 16x16 words: Transformers for image recognition at scale,'' in \emph{International Conference on Learning Representations}, 2021.

\bibitem{gpt1}
A.~Radford, K.~Narasimhan, T.~Salimans, I.~Sutskever \emph{et~al.}, ``Improving language understanding by generative pre-training,'' 2018.

\bibitem{gpt2}
A.~Radford, J.~Wu, R.~Child, D.~Luan, D.~Amodei, I.~Sutskever \emph{et~al.}, ``Language models are unsupervised multitask learners,'' \emph{OpenAI blog}, vol.~1, no.~8, p.~9, 2019.

\bibitem{ernst2004merging}
M.~O. Ernst and H.~H. B{\"u}lthoff, ``Merging the senses into a robust percept,'' \emph{Trends in cognitive sciences}, vol.~8, no.~4, pp. 162--169, 2004.

\bibitem{shams2008benefits}
L.~Shams and A.~R. Seitz, ``Benefits of multisensory learning,'' \emph{Trends in cognitive sciences}, vol.~12, no.~11, pp. 411--417, 2008.

\bibitem{ramachandran2001synaesthesia}
V.~S. Ramachandran and E.~M. Hubbard, ``Synaesthesia--a window into perception, thought and language,'' \emph{Journal of consciousness studies}, vol.~8, no.~12, pp. 3--34, 2001.

\bibitem{cytowic2002synesthesia}
R.~E. Cytowic, \emph{Synesthesia: A union of the senses}.\hskip 1em plus 0.5em minus 0.4em\relax MIT press, 2002.

\bibitem{qwen2audio}
Y.~Chu, J.~Xu, Q.~Yang, H.~Wei, X.~Wei, Z.~Guo, Y.~Leng, Y.~Lv, J.~He, J.~Lin \emph{et~al.}, ``Qwen2-audio technical report,'' \emph{arXiv preprint arXiv:2407.10759}, 2024.

\bibitem{qwen2vl}
P.~Wang, S.~Bai, S.~Tan, S.~Wang, Z.~Fan, J.~Bai, K.~Chen, X.~Liu, J.~Wang, W.~Ge \emph{et~al.}, ``Qwen2-vl: Enhancing vision-language model's perception of the world at any resolution,'' \emph{arXiv preprint arXiv:2409.12191}, 2024.

\bibitem{qwen2.5omni}
J.~Xu, Z.~Guo, J.~He, H.~Hu, T.~He, S.~Bai, K.~Chen, J.~Wang, Y.~Fan, K.~Dang \emph{et~al.}, ``Qwen2. 5-omni technical report,'' \emph{arXiv preprint arXiv:2503.20215}, 2025.

\bibitem{vggsound}
H.~Chen, W.~Xie, A.~Vedaldi, and A.~Zisserman, ``Vggsound: A large-scale audio-visual dataset,'' in \emph{ICASSP 2020-2020 IEEE International Conference on Acoustics, Speech and Signal Processing (ICASSP)}.\hskip 1em plus 0.5em minus 0.4em\relax IEEE, 2020, pp. 721--725.

\bibitem{pann}
Q.~Kong, Y.~Cao, T.~Iqbal, Y.~Wang, W.~Wang, and M.~D. Plumbley, ``Panns: Large-scale pretrained audio neural networks for audio pattern recognition,'' \emph{IEEE/ACM Transactions on Audio, Speech, and Language Processing}, vol.~28, pp. 2880--2894, 2020.

\bibitem{ast}
Y.~Gong, Y.-A. Chung, and J.~Glass, ``Ast: Audio spectrogram transformer,'' in \emph{Interspeech 2021}, 2021, pp. 571--575.

\bibitem{sam}
A.~Kirillov, E.~Mintun, N.~Ravi, H.~Mao, C.~Rolland, L.~Gustafson, T.~Xiao, S.~Whitehead, A.~C. Berg, W.-Y. Lo \emph{et~al.}, ``Segment anything,'' in \emph{Proceedings of the IEEE/CVF international conference on computer vision}, 2023, pp. 4015--4026.

\bibitem{clip}
A.~Radford, J.~W. Kim, C.~Hallacy, A.~Ramesh, G.~Goh, S.~Agarwal, G.~Sastry, A.~Askell, P.~Mishkin, J.~Clark \emph{et~al.}, ``Learning transferable visual models from natural language supervision,'' in \emph{International conference on machine learning}.\hskip 1em plus 0.5em minus 0.4em\relax PmLR, 2021, pp. 8748--8763.

\bibitem{clap}
B.~Elizalde, S.~Deshmukh, M.~Al~Ismail, and H.~Wang, ``Clap learning audio concepts from natural language supervision,'' in \emph{ICASSP 2023-2023 IEEE International Conference on Acoustics, Speech and Signal Processing (ICASSP)}.\hskip 1em plus 0.5em minus 0.4em\relax IEEE, 2023, pp. 1--5.

\bibitem{audiobench}
\BIBentryALTinterwordspacing
B.~Wang, X.~Zou, G.~Lin, S.~Sun, Z.~Liu, W.~Zhang, Z.~Liu, A.~Aw, and N.~F. Chen, ``{A}udio{B}ench: A universal benchmark for audio large language models,'' in \emph{Proceedings of the 2025 Conference of the Nations of the Americas Chapter of the Association for Computational Linguistics: Human Language Technologies (Volume 1: Long Papers)}, L.~Chiruzzo, A.~Ritter, and L.~Wang, Eds.\hskip 1em plus 0.5em minus 0.4em\relax Albuquerque, New Mexico: Association for Computational Linguistics, Apr. 2025, pp. 4297--4316. [Online]. Available: \url{https://aclanthology.org/2025.naacl-long.218/}
\BIBentrySTDinterwordspacing

\bibitem{lee2024vhelm}
T.~Lee, H.~Tu, C.~H. Wong, W.~Zheng, Y.~Zhou, Y.~Mai, J.~Roberts, M.~Yasunaga, H.~Yao, C.~Xie \emph{et~al.}, ``Vhelm: A holistic evaluation of vision language models,'' \emph{Advances in Neural Information Processing Systems}, vol.~37, pp. 140\,632--140\,666, 2024.

\bibitem{avhbench}
K.~Sung-Bin, O.~Hyun-Bin, J.~Lee, A.~Senocak, J.~S. Chung, and T.-H. Oh, ``{AVHB}ench: A cross-modal hallucination benchmark for audio-visual large language models,'' in \emph{The Thirteenth International Conference on Learning Representations}, 2025.

\bibitem{avtrustbench}
S.~Chowdhury, S.~Nag, S.~Dasgupta, Y.~Wang, M.~Elhoseiny, R.~Gao, and D.~Manocha, ``Avtrustbench: Assessing and enhancing reliability and robustness in audio-visual llms,'' \emph{arXiv preprint arXiv:2501.02135}, 2025.

\bibitem{gu2024minillm}
Y.~Gu, L.~Dong, F.~Wei, and M.~Huang, ``Mini{LLM}: Knowledge distillation of large language models,'' in \emph{The Twelfth International Conference on Learning Representations}, 2024.

\bibitem{efficientvlm}
\BIBentryALTinterwordspacing
T.~Wang, W.~Zhou, Y.~Zeng, and X.~Zhang, ``{E}fficient{VLM}: Fast and accurate vision-language models via knowledge distillation and modal-adaptive pruning,'' in \emph{Findings of the Association for Computational Linguistics: ACL 2023}, A.~Rogers, J.~Boyd-Graber, and N.~Okazaki, Eds.\hskip 1em plus 0.5em minus 0.4em\relax Toronto, Canada: Association for Computational Linguistics, Jul. 2023, pp. 13\,899--13\,913. [Online]. Available: \url{https://aclanthology.org/2023.findings-acl.873/}
\BIBentrySTDinterwordspacing

\bibitem{chen2021distilling}
Y.~Chen, Y.~Xian, A.~Koepke, Y.~Shan, and Z.~Akata, ``Distilling audio-visual knowledge by compositional contrastive learning,'' in \emph{Proceedings of the IEEE/CVF conference on computer vision and pattern recognition}, 2021, pp. 7016--7025.

\bibitem{sarkar2024xkd}
P.~Sarkar and A.~Etemad, ``Xkd: Cross-modal knowledge distillation with domain alignment for video representation learning,'' in \emph{Proceedings of the AAAI Conference on Artificial Intelligence}, vol.~38, no.~13, 2024, pp. 14\,875--14\,885.

\bibitem{qwenaudio}
Y.~Chu, J.~Xu, X.~Zhou, Q.~Yang, S.~Zhang, Z.~Yan, C.~Zhou, and J.~Zhou, ``Qwen-audio: Advancing universal audio understanding via unified large-scale audio-language models,'' \emph{arXiv preprint arXiv:2311.07919}, 2023.

\bibitem{kimiaudio}
D.~Ding, Z.~Ju, Y.~Leng, S.~Liu, T.~Liu, Z.~Shang, K.~Shen, W.~Song, X.~Tan, H.~Tang \emph{et~al.}, ``Kimi-audio technical report,'' \emph{arXiv preprint arXiv:2504.18425}, 2025.

\bibitem{qwen2.5vl}
S.~Bai, K.~Chen, X.~Liu, J.~Wang, W.~Ge, S.~Song, K.~Dang, P.~Wang, S.~Wang, J.~Tang \emph{et~al.}, ``Qwen2. 5-vl technical report,'' \emph{arXiv preprint arXiv:2502.13923}, 2025.

\bibitem{videollama3}
B.~Zhang, K.~Li, Z.~Cheng, Z.~Hu, Y.~Yuan, G.~Chen, S.~Leng, Y.~Jiang, H.~Zhang, X.~Li \emph{et~al.}, ``Videollama 3: Frontier multimodal foundation models for image and video understanding,'' \emph{arXiv preprint arXiv:2501.13106}, 2025.

\bibitem{qwen2}
\BIBentryALTinterwordspacing
A.~Yang, B.~Yang, B.~Hui, B.~Zheng, B.~Yu, C.~Zhou, C.~Li, C.~Li, D.~Liu, F.~Huang, G.~Dong, H.~Wei, H.~Lin, J.~Tang, J.~Wang, J.~Yang, J.~Tu, J.~Zhang, J.~Ma, J.~Yang, J.~Xu, J.~Zhou, J.~Bai, J.~He, J.~Lin, K.~Dang, K.~Lu, K.~Chen, K.~Yang, M.~Li, M.~Xue, N.~Ni, P.~Zhang, P.~Wang, R.~Peng, R.~Men, R.~Gao, R.~Lin, S.~Wang, S.~Bai, S.~Tan, T.~Zhu, T.~Li, T.~Liu, W.~Ge, X.~Deng, X.~Zhou, X.~Ren, X.~Zhang, X.~Wei, X.~Ren, X.~Liu, Y.~Fan, Y.~Yao, Y.~Zhang, Y.~Wan, Y.~Chu, Y.~Liu, Z.~Cui, Z.~Zhang, Z.~Guo, and Z.~Fan, ``Qwen2 technical report,'' 2024. [Online]. Available: \url{https://arxiv.org/abs/2407.10671}
\BIBentrySTDinterwordspacing

\bibitem{cl1}
Z.~Wang, C.~Subakan, X.~Jiang, J.~Wu, E.~Tzinis, M.~Ravanelli, and P.~Smaragdis, ``Learning representations for new sound classes with continual self-supervised learning,'' \emph{IEEE Signal Processing Letters}, vol.~29, pp. 2607--2611, 2022.

\bibitem{cl2}
X.~Jiang, Y.~A. Li, and N.~Mesgarani, ``Decor: Defy knowledge forgetting by predicting earlier audio codes,'' in \emph{Interspeech 2023}, 2023, pp. 2818--2822.

\bibitem{lora}
E.~J. Hu, Y.~Shen, P.~Wallis, Z.~Allen-Zhu, Y.~Li, S.~Wang, L.~Wang, W.~Chen \emph{et~al.}, ``Lora: Low-rank adaptation of large language models.'' \emph{ICLR}, vol.~1, no.~2, p.~3, 2022.

\bibitem{audioset}
J.~F. Gemmeke, D.~P. Ellis, D.~Freedman, A.~Jansen, W.~Lawrence, R.~C. Moore, M.~Plakal, and M.~Ritter, ``Audio set: An ontology and human-labeled dataset for audio events,'' in \emph{2017 IEEE international conference on acoustics, speech and signal processing (ICASSP)}.\hskip 1em plus 0.5em minus 0.4em\relax IEEE, 2017, pp. 776--780.

\bibitem{aadllm}
X.~Jiang, S.~S. Dindar, V.~Choudhari, S.~Bickel, A.~Mehta, G.~M. McKhann, D.~Friedman, A.~Flinker, and N.~Mesgarani, ``Aad-llm: Neural attention-driven auditory scene understanding,'' \emph{arXiv preprint arXiv:2502.16794}, 2025.

\bibitem{sood2023multimodal}
E.~Sood, F.~K{\"o}gel, P.~M{\"u}ller, D.~Thomas, M.~B{\^a}ce, and A.~Bulling, ``Multimodal integration of human-like attention in visual question answering,'' in \emph{Proceedings of the IEEE/CVF Conference on Computer Vision and Pattern Recognition}, 2023, pp. 2648--2658.

\end{thebibliography}







\end{document}